\documentclass[prl,aps,twocolumn,superscriptaddress,showpacs]{revtex4-1}

\usepackage[dvips]{graphicx} 
\usepackage{amsfonts}
\usepackage{amssymb}
\usepackage{amscd}
\usepackage{amsmath}    
\usepackage{enumerate}
\usepackage{epsfig}
\usepackage{subfigure}

\begin{document}

\title{Measurement-device-independent quantum key distribution over 200 km}

\author{Yan-Lin Tang}
\author{Hua-Lei Yin}
\affiliation{Department of Modern Physics and National Laboratory for Physical Sciences at Microscale, Shanghai Branch, University of Science and Technology of China, Hefei, Anhui 230026, China}
\affiliation{CAS Center for Excellence and Synergetic Innovation Center in Quantum Information and Quantum Physics, Shanghai Branch,  University of Science and Technology of China, Hefei, Anhui 230026, China}
\author{Si-Jing Chen}
\affiliation{State Key Laboratory of Functional Materials for Informatics, Shanghai Institute of Microsystem and Information Technology, Chinese Academy of Sciences, Shanghai 200050, China}
\author{Yang Liu}
\affiliation{Department of Modern Physics and National Laboratory for Physical Sciences at Microscale, Shanghai Branch, University of Science and Technology of China, Hefei, Anhui 230026, China}
\affiliation{CAS Center for Excellence and Synergetic Innovation Center in Quantum Information and Quantum Physics, Shanghai Branch,  University of Science and Technology of China, Hefei, Anhui 230026, China}
\author{Wei-Jun Zhang}
\affiliation{State Key Laboratory of Functional Materials for Informatics, Shanghai Institute of Microsystem and Information Technology, Chinese Academy of Sciences, Shanghai 200050, China}
\author{Xiao Jiang}
\affiliation{Department of Modern Physics and National Laboratory for Physical Sciences at Microscale, Shanghai Branch, University of Science and Technology of China, Hefei, Anhui 230026, China}
\affiliation{CAS Center for Excellence and Synergetic Innovation Center in Quantum Information and Quantum Physics, Shanghai Branch,  University of Science and Technology of China, Hefei, Anhui 230026, China}
\author{Lu Zhang}
\affiliation{State Key Laboratory of Functional Materials for Informatics, Shanghai Institute of Microsystem and Information Technology, Chinese Academy of Sciences, Shanghai 200050, China}
\author{Jian Wang}
\affiliation{Department of Modern Physics and National Laboratory for Physical Sciences at Microscale, Shanghai Branch, University of Science and Technology of China, Hefei, Anhui 230026, China}
\affiliation{CAS Center for Excellence and Synergetic Innovation Center in Quantum Information and Quantum Physics, Shanghai Branch,  University of Science and Technology of China, Hefei, Anhui 230026, China}
\author{Li-Xing You}
\affiliation{State Key Laboratory of Functional Materials for Informatics, Shanghai Institute of Microsystem and Information Technology, Chinese Academy of Sciences, Shanghai 200050, China}
\author{Jian-Yu Guan}
\author{Dong-Xu Yang}
\affiliation{Department of Modern Physics and National Laboratory for Physical Sciences at Microscale, Shanghai Branch, University of Science and Technology of China, Hefei, Anhui 230026, China}
\affiliation{CAS Center for Excellence and Synergetic Innovation Center in Quantum Information and Quantum Physics, Shanghai Branch,  University of Science and Technology of China, Hefei, Anhui 230026, China}
\author{Zhen Wang}
\affiliation{State Key Laboratory of Functional Materials for Informatics, Shanghai Institute of Microsystem and Information Technology, Chinese Academy of Sciences, Shanghai 200050, China}
\author{Hao Liang}
\affiliation{Department of Modern Physics and National Laboratory for Physical Sciences at Microscale, Shanghai Branch, University of Science and Technology of China, Hefei, Anhui 230026, China}
\affiliation{CAS Center for Excellence and Synergetic Innovation Center in Quantum Information and Quantum Physics, Shanghai Branch,  University of Science and Technology of China, Hefei, Anhui 230026, China}
\author{Zhen Zhang}
\affiliation{Center for Quantum Information, Institute for Interdisciplinary Information Sciences, Tsinghua University, Beijing, 100084, China}
\affiliation{CAS Center for Excellence and Synergetic Innovation Center in Quantum Information and Quantum Physics, Shanghai Branch,  University of Science and Technology of China, Hefei, Anhui 230026, China}
\author{Nan Zhou}
\affiliation{Department of Modern Physics and National Laboratory for Physical Sciences at Microscale, Shanghai Branch, University of Science and Technology of China, Hefei, Anhui 230026, China}
\affiliation{CAS Center for Excellence and Synergetic Innovation Center in Quantum Information and Quantum Physics, Shanghai Branch,  University of Science and Technology of China, Hefei, Anhui 230026, China}
\author{Xiongfeng Ma}
\affiliation{Center for Quantum Information, Institute for Interdisciplinary Information Sciences, Tsinghua University, Beijing, 100084, China}
\affiliation{CAS Center for Excellence and Synergetic Innovation Center in Quantum Information and Quantum Physics, Shanghai Branch,  University of Science and Technology of China, Hefei, Anhui 230026, China}
\author{Teng-Yun Chen}
\author{Qiang Zhang}
\author{Jian-Wei Pan}
\affiliation{Department of Modern Physics and National Laboratory for Physical Sciences at Microscale, Shanghai Branch, University of Science and Technology of China, Hefei, Anhui 230026, China}
\affiliation{CAS Center for Excellence and Synergetic Innovation Center in Quantum Information and Quantum Physics, Shanghai Branch, University of Science and Technology of China, Hefei, Anhui 230026, China}

\begin{abstract}
Measurement-device-independent quantum key distribution (MDIQKD) protocol is immune to all attacks on detection and guarantees the information-theoretical security even with imperfect single photon detectors. Recently, several proof-of-principle demonstrations of MDIQKD have been achieved. Those experiments, although novel, are implemented through limited distance with a key rate less than 0.1 bps. Here, by developing a 75 MHz clock rate fully-automatic and highly-stable system, and superconducting nanowire single photon detectors with detection efficiencies more than 40\%, we extend the secure transmission distance of MDIQKD to 200 km and achieve a secure key rate of three orders of magnitude higher. These results pave the way towards a quantum network with measurement-device-independent security.

\end{abstract}


\maketitle

Quantum key distribution (QKD) \cite{Bennett:BB84:1984,Ekert:QKD:1991} offers the most appealing solution for secure key exchange by providing information-theoretical security. Despite the tremendous experimental developments, practical QKD systems still suffer from various attacks rooted in their deviations from the theoretical models in security proofs.
The security of the measurement-device-independent quantum key distribution (MDIQKD) protocol \cite{Lo:MIQKD:2012}, inspired by the time-reversed EPR-based QKD protocol \cite{Biham:1996:Quantum,Inamori:TimeReverseEPR:2002}, does not rely on any assumption on measurement. Compared with the regular QKD system where Eve can take advantage of the side information of the detection devices, the MDIQKD protocol allows Eve to have a full control of the measurement devices without any key information leakage. Thus it is naturally immune to all detection attacks \cite{MAS_Eff_06,Qi:TimeShift:2007,Lydersen:Hacking:2010,Weier:DeadtimeAttack:2011,Mark:DamageAtt:2014}, which are believed as the main threat to practical QKD systems.
Meanwhile, the MDIQKD protocol not only provides security with imperfect detectors, but also is able to achieve performance comparable to the regular prepare-and-measure QKD systems \cite{Hiroki:QKD40dB:2007,liu:decoy:2010,Gisin:QKD250km:2009,Guo:DPS260km:2012}.

Due to its security and practicability, the MDIQKD protocol has attracted extensive attentions in the field.
Up till now, many efforts have been devoted to experimental demonstrations of the MDIQKD protocol using time-bin phase encoding scheme \cite{Tittel:MDIQKDFielfTest:2013,Liu:MIQKDexp:2013} and polarization encoding scheme \cite{Silva:DemoPolMDIQKD:2013,tang:experimental:2013}. Previously, two full implementations of the MDIQKD protocol, which employ the decoy-state method \cite{Hwang:Decoy:2003,Lo:Decoy:2005,Wang:Decoy:2005} to guarantee the security of source, achieved the secure key rates of $0.12 \ bps$ \cite{Liu:MIQKDexp:2013} and $0.0047 \ bps$ \cite{tang:experimental:2013} over 50 km and 10 km fiber links, respectively.
These experimental demonstrations have formed a solid basis for the feasibility of MDIQKD systems. However, compared with standard QKD systems, these demonstrations have limited transmission distances and low key rates. The demonstration of the practicability of the MDIQKD protocol is still missing.


In this Letter, by improving the system clock rate, single photon detector efficiency, and stability, we achieve the secure transmission distance to 200 km, and improve the secure key rate by three orders of magnitude higher than previous demonstrations.

\begin{figure*}[tbh]
\centering
\resizebox{12cm}{!}{\includegraphics{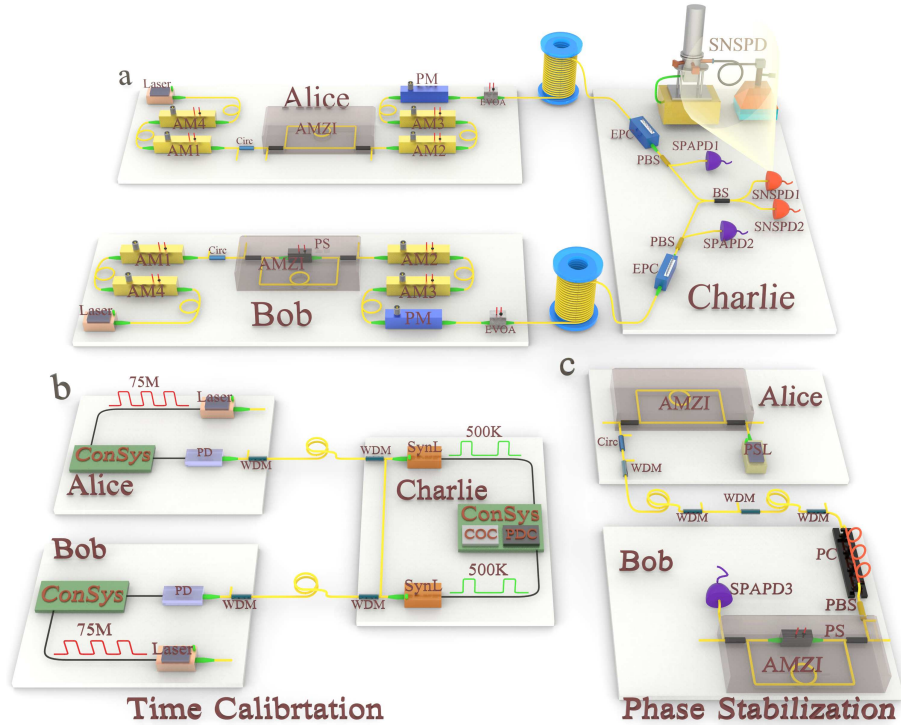}}
\caption{(a) Schematic layout of our MDIQKD setup. Alice's (Bob's) signal laser pulses (1550 nm) are modulated into three decoy-state intensities by AM1.
An AMZI, AM2$\sim$4 and one PM are to encode qubits.
Charlie's setup consists of a polarization stabilization system and a BSM system. The polarization stabilization system in each link includes an EPC, a PBS and a SPAPD. The BSM system includes an interference BS and two SNSPDs.
(b) Time calibration system. Two SynLs (1570 nm) are adopted, with the 500 kHz shared time reference generated from a crystal oscillator circuit (COC) and with the time delayed by a programmable delay chip (PDC). Alice (Bob) receives the SynL pulses with a PD and then regenerates a system clock of 75 MHz. WDM: wavelength division multiplexer, ConSys: control system.
(c) Phase stabilization system. Circ: circulator, PC: polarization controller, PS: phase shifter.}
\label{Fig:LabMDIQKD}
\end{figure*}

The MDIQKD experiment setup is illustrated in Fig.~\ref{Fig:LabMDIQKD}(a), where one can see that the positions of Alice and Bob are symmetric. Alice's (Bob's) signal laser source (1550 nm) is internally modulated into a pulse train with a width of 2.5 ns and a clock rate of 75 MHz, compared with 1 MHz in previous experiments \cite{Tittel:MDIQKDFielfTest:2013,Liu:MIQKDexp:2013,Silva:DemoPolMDIQKD:2013,tang:experimental:2013}. The directly modulated pulse trains' phases are intrinsically random to make sure the system is immune to the unambiguous-state-discrimination attack \cite{tang:USDAttack:2012}. Alice (Bob) uses an amplitude modulator (AM) and electrical variable optical attenuator (EVOA) to randomly modulate the laser into three different intensities according to the decoy-state method, one as signal state intensity ($\mu=0.4$), another as decoy state intensity ($\nu=0.07$), and the rest as the vacuum state intensity ($0$). Their probability distribution is set as $33\%$, $45\%$ and $22\%$. The setting of the intensity and probability distribution is optimized at the distance of 200 km, and is adopted for all the distances in our experiment.

We employ the time-bin phase-encoding scheme and utilize a combination of an asymmetrical Mach-Zehnder interferometer (AMZI),
three AMs and one phase modulator (PM) to encode qubits. The AMZI divides the laser pulse into two time bins separated by 6.5 ns time delay.
For the $Z$ basis, the key bit is encoded in time bin, 0 or 1, by AM2 and AM3. For the $X$ basis, the key bit is encoded into the relative phase, 0 or $\pi$, by PM. The average photon number is controlled by AM4. All the modulators, including the AMs and the PM, are controlled by the random numbers of Alice and Bob independently. We remark that in order to increase the fidelity of time bin 0 or 1, we exploit two AMs for the $Z$-basis encoding. Beneficially, this arrangement helps to improve the extinction ratio of the vacuum state in the decoy-state method.

The laser pulses of Alice (Bob) go through a fiber spool of the length ranging from 25 km to 100 km in each arm, to interfere with the ones sent by Bob (Alice). In the middle, Charlie takes a partial Bell state measurement (BSM). The critical challenge for our system is to develop a stable BSM system for two independent laser pulses traveling through two 100-km-fiber links, under a high clock rate. Note that it is not a trivial upgrade comparing the previous MDIQKD systems, because under a high clock rate, the task imposes a technical challenge on rigorous timing and frequency calibration. Furthermore, the time and polarization drifting due to the 200 km channel adds further challenge to the experiment.



In our setup, we develop several automatic feedback systems to calibrate the time, spectrum and polarization modes of the two independent laser pulses.

For the timing mode shown in Fig.~\ref{Fig:LabMDIQKD}(b), two synchronization laser (SynL, 1570 nm) pulse trains are sent through two additional fiber links from Charlie to Alice and Bob, with shared time references generated by a crystal oscillator circuit at Charlie's site.
Alice (Bob) utilizes a photoelectric detector (PD) to detect the SynL pulses. The output signals of the PD are used to regenerate a 75 MHz system clock, so that the whole system can be synchronized.
Then we precisely overlap the two signal laser pulse trains via a feedback control. Alice and Bob alternatively send the signal laser pulses to Charlie. Charlie uses the SNSPD to measure the arriving time of the signal laser pulses. Based on the arriving time difference, Charlie adjusts the time delay between the two SynL pulse trains with a programmable delay chip. The timing resolution is 10 ps and the total timing calibration precision is below 20 ps, both of which are much smaller than the 2.5 ns pulse width of the signal laser.

For the spectrum mode, we utilize an optical spectrum analyzer and a temperature controlled circuit built in the laser, to acquire and then compensate the central wavelength difference of Alice's and Bob's lasers.
We first select two nearly identical laser diodes as Alice's and Bob's laser sources, considering the aspects of both the same full width at half maximum (FWHM) wavelength and the same central wavelength. Then we utilize an optical spectrum analyzer to measure the central wavelength of Alice's and Bob's lasers, respectively. At last, we set the two laser wavelengths as the same by adjusting their temperatures through the temperature controlled circuits built in the lasers. The controlling precision of the central wavelength is about 0.5 pm, which is small compared with the 16 pm FWHM and promises an indistinguishable overlap of Alice's and Bob's spectrum modes.

For the polarization mode, we adopt a polarization stabilization system comprised of an electric polarization controller (EPC), a polarization beam splitter (PBS), and an InGaAs/InP single-photon avalanche photodiode (SPAPD).
We insert the EPC and PBS before the interference beam splitter (BS), and connect the transmission port of PBS with the BS. The reflection port of the PBS is monitored by the SPAPD, whose count rate is used as the feedback signal to control the EPC.
Using this polarization stabilization system operated in real time, the polarization mode is maintained and the fluctuation of received laser power in Charlie's site is controlled to be less than 3\%.

Besides these calibration systems above, we adopt a phase stabilization system \cite{Liu:MIQKDexp:2013} to maintain the phase reference frames of Alice's and Bob's phase-encoding setups, as shown in Fig.~\ref{Fig:LabMDIQKD}(c).
The phase reference frame, namely the relative phase between AMZI's two arms, may fluctuate with temperature and stress, and the fluctuation will introduce further errors in the $X$ basis. To stabilize this, we employ a phase-stabilization laser (PSL, 1550 nm). Alice sends the laser pulses of PSL through Alice's and Bob's AMZIs connected by an additional fiber between Alice and Bob. Bob monitors the power at an output of his AMZI with another InGaAs/InP SPAPD.
The phase is then calibrated by a phase shifter inside Bob's AMZI. Besides, we put the AMZIs in a thermal container to isolate the temperature and stress perturbation.

All the aforementioned feedback systems contribute to a good interference and high stabilization, and the automatic calibration procedure can largely improve the time utilization efficiency. While the previous experiment \cite{Liu:MIQKDexp:2013} needs to manually calibrate the interference per 15 minutes, the current setup can remain working for more than one day.

In Charlie's site, a partial BSM is implemented with an interference BS and two superconducting nanowire single photon detectors (SNSPDs). The insertion loss of the measurement system is 1 dB. A Bell state is post-selected when the two detectors in the two output arms of the BS have a coincidence at two alternative
time bins, i.e., the first detector has an event at time bin 0 (1) and simultaneously the second detector has an event at time bin 1 (0).
Since the key is generated based on two detectors' coincidence measurement, MDIQKD has a higher requirement for the detector efficiency
compared with regular QKD scheme.  
In the experiment, we develop a multi-channel SNSPD system cooled by a Gifford-McMahon cryocooler. The SNSPDs are fabricated from ultrathin NbN film on SiO$_2$/Si substrate with a typical meandered nanowire structure and an optical cavity structure. No bandpass filter is included \cite{Yang:SNSPD:2014}.
Operated at 2.2 K, two SNSPDs provide the system detection efficiencies of 46\% and 40\% at the dark count rate of 10 Hz, respectively, which are almost two or three times higher than the efficiency in previous experiments.
In order to achieve a better signal noise ratio, we adopt a time window of 1.5 ns, 60\% of the pulse width of 2.5 ns.


After Charlie announcing the BSM results, Alice and Bob then sift out a raw key stream, and generate the final secure key through error correction and privacy amplification, detailed in Section I of Supplemental Material.
We assume that the secure key is extracted from the data when both Alice and Bob encode their pulses using signal states in the $Z$ basis. The secure key rate formula is \cite{Lo:Decoy:2005,Lo:MIQKD:2012},
\begin{equation} \label{MIExp:Post:KeyrateMI}
\begin{aligned}
R &\ge Q_{11}^{\mu\mu}[1-H(e_{11}^{\mu\mu})] - Q^{\mu\mu} f H(E^{\mu\mu}),
\end{aligned}
\end{equation}
where $Q^{\mu\mu}$ and $E^{\mu\mu}$ are the overall gain and error rate when both sources generate signal states. $Q_{11}^{\mu\mu}$ and $e_{11}^{\mu\mu}$, the gain and phase error rate when both sources generate single-photon states within signal states, can be estimated by the decoy-state method. The parameter $f$ is the error correction efficiency, and we take the value $f=1.16$ in our calculation. $H(e)=-e\log_{2}(e)-(1-e)\log_{2}(1-e)$ is the binary Shannon entropy function.

We have continuously run the system in the laboratory for 130 hours with spooled fibers of distances of 50 km, 100 km, 150 km and 200 km. The detail for the experimental results can be found in Section II of Supplemental Material. To calculate the secure key rate, we use the decoy-state method and make the finite-key analysis with the Chernoff bound to estimate $Q_{11}^{\mu\mu}$ and $e_{11}^{\mu\mu}$, detailed in Section I of Supplemental Material. As shown in Fig.~\ref{Fig:Results}, the secure key rates are marked, which fit the theoretical curve well. Previous experimental results \cite{Liu:MIQKDexp:2013,tang:experimental:2013} are also shown in Fig.~\ref{Fig:Results} for comparison. One can see that we have improved the secure key rate by three orders of magnitude than the previous results \cite{Liu:MIQKDexp:2013,tang:experimental:2013}.

\begin{figure}[tbh]
\centering
\resizebox{8.6cm}{!}{\includegraphics{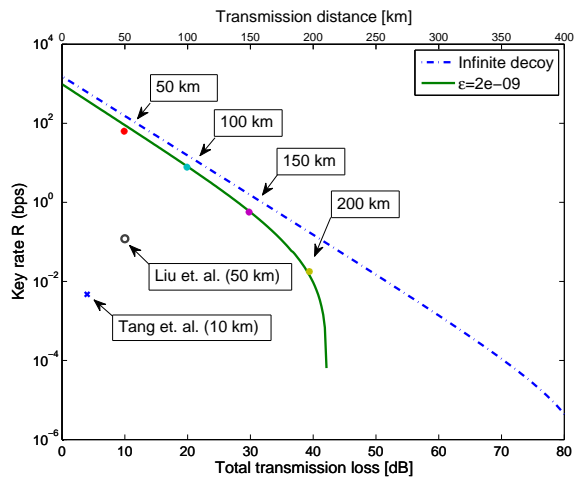}}
\caption{Secure key rates of experiments in the laboratory, as well as the simulation results. The four dots correspond to the experimental results with the fiber transmitting loss of 9.9 dB (50 km), 19.9 dB (100 km), 29.8 dB (150 km) and 39.6 dB (200 km). The solid curve shows the result calculated by simulating the vacuum+weak decoy state scheme with the experimental parameters. The dashed curve represents the optimal result with infinite number of decoy states.}
\label{Fig:Results}
\end{figure}

Taking the experiment result of the 200 km detailed in the Supplemental Material for example, we can see that $E^{\mu\nu}$ $(\mu,\nu \neq 0)$ of the $Z$ basis is less than 0.25\%, owing to the fact that the two AMs employed to encode the $Z$-basis qubits contribute to an extinction ratio more than 40 dB.  Meanwhile, $E^{\mu\mu}$ of the $X$ basis is less than 26\% compared with the typical value 25\%, from which we can infer that the interference of Alice's and Bob's laser pulse is good and the automatic feedback systems operate effectively.  Through error correction and privacy amplification, the final secure key rate obtained is $0.009\ bps$.

With this MDIQKD system, we have extended the distribution distance from 50 km to 200 km, and filled the gap of attainable distance between the MDIQKD protocol and the regular BB84 protocol. In addition, the secure key rate is higher than the previous results by three orders of magnitude. These results demonstrate the practicability of MDIQKD. Furthermore, we have moved the system into installed fiber network and implemented a field test \cite{our field test}.

We remark that the techniques developed in our MDIQKD system pave the way for other quantum communication tasks, such as quantum repeater \cite{Zoller:QRepeater:1998} and quantum network \cite{NNews:Leapout:2014}. The MDIQKD protocol has an intrinsic property which is desirable for constructing quantum network with the star-type structure, since the detection system placed in the BSM site (as a server) can be shared by all the transmitters. To add more transmitters in the network, we only need the laser sources and the modulators which are much cheaper and smaller than the detection system. We can expect that the MDIQKD network can be built within reach of current technology and become mature in the near future. Especially, since a global-scale QKD system using communication satellite \cite{Zeilinger:144kmQKD:2007} should tolerate around 35 dB channel loss, our results have covered this tolerance range in our experiment. We believe that the MDIQKD protocol is a good choice not only for terrestrial QKD over deployed fibers, but also for the satellite-based global QKD.


We remark that since the clock rate is mainly restricted by the overall timing jitter, 10 GHz clock rate is achievable with the state-of-the-art components \cite{Hiroki:QKD40dB:2007}. Besides, there is still much room for the further improvement of SNSPD efficiency \cite{marsili:SSPD93:2013}. We can extrapolate that the transmission distance and secure key rate can be further improved by increasing the clock rate and detector efficiency.


\textbf{\subsection*{Acknowledgments}}
The authors would like to thank Xiaoming Xie and Mianheng Jiang for enlightening discussions, especially to Chunli Luo for her experimental assist. This work has been supported by the National Fundamental Research Program (under Grant No. 2011CB921300, 2013CB336800 and 2011CBA00300), the National Natural Science Foundation of China, the Chinese Academy of Science, and the Shandong Institute of Quantum Science \& Technology Co., Ltd.






\end{document}